# A Message-Passing Algorithm for Counting Short Cycles in a Graph

Mehdi Karimi, *Student Member, IEEE,* Amir H. Banihashemi, *Senior Member, IEEE*

Department of Systems and Computer Engineering, Carleton University,

Ottawa, Ontario, Canada

Email: mkarimi, ahashemi@sce.carleton.ca

## Abstract

A message-passing algorithm for counting short cycles in a graph is presented. For bipartite graphs, which are of particular interest in coding, the algorithm is capable of counting cycles of length $g, g + 2, \ldots, 2g - 2$, where $g$ is the girth of the graph. For a general (non-bipartite) graph, cycles of length $g, g + 1, \ldots, 2g - 1$ can be counted. The algorithm is based on performing integer additions and subtractions in the nodes of the graph and passing extrinsic messages to adjacent nodes. The complexity of the proposed algorithm grows as $O(g|E|^2)$, where $|E|$ is the number of edges in the graph. For sparse graphs, the proposed algorithm significantly outperforms the existing algorithms in terms of computational complexity and memory requirements.

## I. INTRODUCTION

Graphical models are widely used in different branches of science and engineering to represent systems and facilitate the description of inference algorithms. The structure of the graphs consequently plays an important role in the dynamics of the system and the performance of the corresponding algorithms. One important example, which has many applications in areas such as artificial intelligence, signal processing and digital communications, is the *factor graph* representation of systems and the *sum-product* algorithm [10]. Factor graphs are bipartite graphs and the sum-product algorithm is a generic *message-passing* algorithm which operates in a factor graph. One notable application of factor graphs and message-passing algorithms is in channel coding, where widely popular schemes such as *turbo codes* [3] and *low-density parity-check (LDPC) codes* [7] can be considered as specific instances. In particular, a specific instance of a factor graph is a *Tanner graph* [14], which is used to represent an LDPC code. In fact, LDPC codes, which are famous for their capacity-approaching performance on many communication channels, owe their popularity to the good performance of the iterative message-passing algorithms that can decode these codes with relatively low complexity. The low complexity is a consequence of the sparsity of the Tanner graph.

In practical error correction schemes, finite-length codes have to be used. For such codes, the performance of the message-passing algorithms is closely related to the structure of the graph, in general, and its cycles, in particular. In [11], the girth distribution of the Tanner graph was related to the performance of an LDPC code. Numerous publications since have used the cycle structure of the Tanner graph as an important measure of performance of LDPC codes, with the general belief that for good performance, short cycles should be avoided in the Tanner graph of the code. In [9], the authors devised a code construction, known as progressive edge growth (PEG), to maximize the local girth of the graph in a greedy fashion. Halford and Chugg [8] showed that in addition to the girth, the number and statistics of short cycles are also important performance metrics of the code. In [15], error rates of finite-length LDPC codes were accurately and efficiently estimated by enumerating and testing the subsets of short cycles as error patterns. More recently, Asvadi *et al.* [2] devised cyclic liftings that improve the error floor performance of LDPC codes significantly by breaking up the short cycles involved in the dominant trapping sets of the base code. The close relationship between the performance of graph-based coding schemes and the cycle structure of the graph, especially the number of short cycles, motivates the search for efficient algorithms that can count cycles of different length in the graph. In the context of coding, the graph is often bipartite. This includes the Tanner graph of LDPC codes.

Counting the number of cycles in a general graph is known to be a hard problem [6]. Alon *et al.* [1] presented methods for counting short cycles in a general graph. The complexity of their algorithm however is prohibitively high for longer cycles, say beyond 7. Fan and Xiao [5] presented a method for counting cycles of length $2k, 2 \leq k \leq 5$ in the Tanner graph of LDPC codes. The complexity of their method is $O(m^{k+1})$ where $m$ is the number of the check nodes in the graph. Their method quickly becomes prohibitively complex even for counting cycles as short as 6, particularly in graphs with large $m$. An algorithm with similar complexity was proposed in [4] for counting only the shortest cycles of a Tanner graph. Halford and Chugg [8] presented a method for counting short cycles of length $g, g + 2$ and $g + 4$ in bipartite graphs with girth $g$. The complexity of their method is $O(gn^3)$, where $n$ is the size of the larger set between the two node partitions.

In this paper, we present an algorithm that counts the cycles of length $g, g+2, \ldots, 2g-2$ in a bipartite graph. The algorithm is based on message-passing on the edges of the graph, where the messages are computed at the nodes with integer additions and subtractions. The algorithm can also be applied to general (non-bipartite) graphs to count cycles of length $g, g+1, \ldots, 2g-1$. The complexity of the proposed algorithm is $O(g|E|^2)$, where $|E|$ is the number of edges in the graph. For sparse bipartite graphs, the proposed algorithm can significantly outperform the algorithm of [8] in terms of both computational complexity and memory requirements. As an example, for a regular graph with node degrees 3 and 6 corresponding to an (8000,4000) LDPC code, the proposed algorithm is more than 30 times faster than the method of [8] and requires less memory by a factor of about 600. Conceptually also, the proposed algorithm is much simpler than the algorithm of [8], in which tedious matrix equations are involved in the counting process. Noteworthy is also the fact that for graphs with $g \geq 6$, the proposed algorithm is capable of counting short cycles of lengths up to at least the same value as the algorithm of [8] does.

The remainder of this paper is organized as follows. Basic definitions and notations are provided in Section II. In Section III, we develop the proposed algorithm and give a simple example. In our presentation, we use bipartite graphs for the sake of simplicity and for the reason that the graphs involved in most coding applications are bipartite. The pseudo code for the algorithm is presented in Section IV. Discussions on complexity and memory requirements and comparisons with the algorithm of [8] will follow in Section V. Section VI contains numerical results. Section VII concludes the paper.

## II. DEFINITIONS AND NOTATIONS

An undirected Graph $G = (V, E)$ is defined as a set of nodes $V$ and a set of edges $E$, where $E$ is some subset of the pairs $\{\{u, v\} : u, v \in V, u \neq v\}$. In this definition and without loss of generality in the context of this paper, we exclude loops using the condition $u \neq v$. Parallel edges are also indistinguishable by this definition and are excluded for simplicity. A *walk* of length $k$ in $G$ is a sequence of nodes $v_1, v_2, \ldots, v_{k+1}$ in $V$ such that $\{v_i, v_{i+1}\} \in E$ for all $i \in \{1, \ldots, k\}$. Equivalently, a walk of length $k$ can be described by the corresponding sequence of $k$ edges. A walk is a *path* if all the nodes $v_1, v_2, \ldots, v_k$ are distinct. A walk is called *closed* if the two end nodes are identical, i.e., if $v_1 = v_{k+1}$ in the previous description. A *cycle* of length $k$ is a closed path of length $k$. In a graph $G$, cycles of length $k$, also referred to as *$k$-cycles*, are denoted by $\mathcal{C}_k$. We use $N_k$ for $|\mathcal{C}_k|$. A closed walk is referred to as *lollipop-style* if no two consecutive edges of the walk are identical. By definition, a lollipop-style closed walk contains at least one cycle. To each (undirected) walk (cycle), we associate two directed walks (cycles), depending on which end node or edge is selected as the starting point. This concept is important in the description of the proposed algorithm since the direction of edges is of consequence in message-passing algorithms.

A graph $G(V, E)$ is called *bipartite* if the set $V$ can be partitioned into two disjoint subsets $U$ and $W$ ($V = U \cup W$ and $U \cap W = \emptyset$) such that every edge in $E$ connects a node from $U$ to a node from $W$. We denote $|U|$ by $n$ and $|W|$ by $m$. Tanner graphs of LDPC codes are bipartite graphs, in which $U$ and $W$ are referred to as *variable nodes* and *check nodes*, respectively. Parameters $n$ and $m$ in this case are the code block length and the number of parity check equations, respectively.

The *girth* $g$ of a graph is the length of a shortest cycle in the graph. For bipartite graphs, all cycles have even lengths and $g$ is an even number. The number of edges connected to a node $v$ is called the *degree* of $v$, and is denoted by $d_v$. We call a bipartite graph $G = (U \cup W, E)$ *regular* if all the nodes in $U$ have the same degree $d_u$ and all the nodes in $W$ have the same degree $d_w$. Otherwise, the graph is called *irregular*. For a regular graph, it is easy to see $n d_u = m d_w = |E|$.

## III. Main Ideas

### A. *Message Passing*

A message-passing algorithm operates in a graph by computing messages at the nodes and passing them along the edges to the adjacent nodes. A well-known example is the sum-product algorithm operating in a factor graph [10]. Message passing algorithms often have the property that a message sent along an edge $e$ is not a function of the message previously received along $e$. We refer to this property as *extrinsic* message-passing. An example is shown in Fig. 1, where the operation at node $v_1$ is multiplication. Extrinsic message-passing, for example, is known to be an important property of good iterative decoders [13]. The algorithm proposed in this paper also has this property.

For bipartite graphs $G(U \cup W, E)$, a natural message-passing schedule is for every node in $U$ to send messages to adjacent nodes in $W$ followed by every node in $W$ to send messages to adjacent nodes in $U$. This is referred to as *parallel schedule* and is used often in iterative decoding algorithms. In this case, a complete cycle of message-passing from $U$ to $W$ and then from $W$ to $U$ is called one *iteration*. We assign discrete time $t$ to message-passing, starting from time index zero followed by positive integer values. Corresponding to a time index $t \geq 0$, we associate an iteration number $\ell = \lfloor t/2 \rfloor + 1 \geq 1$. The time indices $t = 2\ell - 2$ and $t = 2\ell - 1$ correspond to the first and the second halves of the iteration $\ell$. We also refer to messages passed at $t = 0$ as *initial messages*, and use the notation $m_{u \to w}^{(\ell)}$ for a message passed from node $u$ to node $w$ at iteration $\ell$. The notations $m_{\underset{\leftarrow}{u}e}^{(\ell)}$ and $m_{\underset{\rightarrow}{u}e}^{(\ell)}$ are used for the incoming and the outgoing messages to and from node $u$ along edge $e$ at iteration $\ell$, respectively.

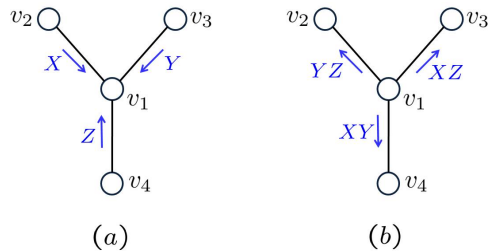

Fig. 1. An extrinsic message-passing algorithm: a) messages received by $v_1$ at $t$, b) messages sent by $v_1$ at $t+1$

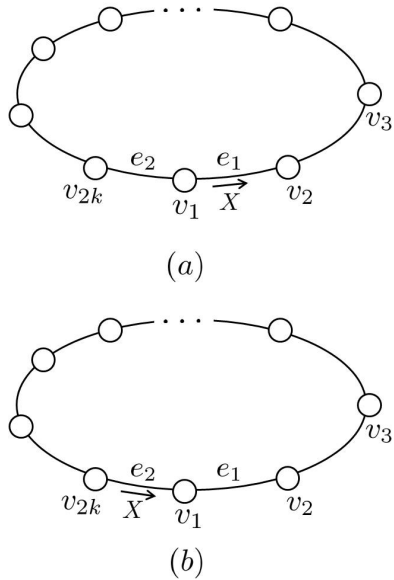

Fig. 2. Message passing for a cycle of length $2k$. a) initial message $X$ is passed along $e_1$, b) after $k$ iterations, $v_1$ receives $X$ along $e_2$.

In the general context of iterative decoding, all nodes in the same partition ($U$ or $W$) perform the same type of operation to generate their messages. The types of operation however are usually different for the two partitions and depend on the nature of the algorithm and the domain in which the messages are presented. In the algorithm developed in this paper, however, all the nodes perform the same type of operation. The messages are all monomials and the operation is multiplication. An example can be seen in Fig. 1. In this work, a monomial is the product of integer powers of variables. For example, a message $m = X_1^i X_2^j X_3^k$ is a monomial with variables $X_1$, $X_2$ and $X_3$. We say $m$ contains $i$ copies of $X_1$, $j$ copies of $X_2$ and $k$ copies of $X_3$. If the variables are ordered, we may use a simpler representation of $m$ as a vector: $m = (i, j, k)$. Using the vector representation of messages, the multiplication of monomials is reduced to the addition of the corresponding vectors.

## B. Algorithm Development

Consider an extrinsic message-passing algorithm in a graph with messages as monomials and node operations as monomial multiplication. In the following, we explain how such an algorithm can count short cycles of the graph. Consider a cycle $C$ of length $2k$ as depicted in Fig. 2(a). Suppose that node $v_1$ of $C$ passes the monomial $X$ as the initial message at $t = 0$ to $v_2$. Due to the extrinsic property of message-passing, $X$ will be passed to $v_3$ from $v_2$ at $t = 1$ and continues its journey around the cycle, one node at a time, until it reaches back to $v_1$ at $t = 2k - 1$ and at the end of iteration $k$, as shown in Fig. 2(b). Clearly, if node $v_1$ had also passed a monomial $Y$ along the edge $e_2$ to $v_{2k}$ at $t = 0$, it would have also received $Y$ from $v_2$ along $e_1$ at the end of iteration $k$. So the iteration number at which node $v_1$ receives back the messages it passed at the first iteration is half the length of the cycle. The following lemma puts this basic idea in the context of the message-passing in a general graph.

*Lemma 1:* Suppose that $C$ is a cycle of length $2k$ in a bipartite graph $G = (V, E)$, and $v \in V$ is in $C$. Denote the two adjacent edges of $v$ in $C$ by $e_1$ and $e_2$. Assume that the message-passing algorithm is initiated on the side of the graph which includes $v$ by passing 1 along every edge in $E$, except $e_1$ and $e_2$. For $e_1$ and $e_2$, the initial messages are monomials $X_1$ and $X_2$, respectively. Then, at iteration $k$, node $v$ will receive one copy of $X_2$ and one copy of $X_1$ along $e_1$ and $e_2$, respectively, where both copies have traveled through all the edges of $C$.

*Proof:* The proof is straight forward and follows directly from the definition of extrinsic message-passing. ∎

It is easy to see that if the node $v$ in Lemma 1 is in $N_{2k}^{v;e_1,e_2}$ cycles of length $2k$ which all include $e_1$ and $e_2$, then at iteration $k$, node $v$ will receive $N_{2k}^{v;e_1,e_2}$ copies of $X_2$ and $N_{2k}^{v;e_1,e_2}$ copies of $X_1$ along $e_1$ and $e_2$, respectively, where each pair of copies has traveled through all the edges of one of the cycles, respectively. Assuming there are no additional copies of $X_2$ received by $v$ along $e_1$ and no additional copies of $X_1$ received by $v$ along $e_2$ at iteration $k$, the monomials received at iteration $k$ by $v$ along $e_1$ and $e_2$ are respectively $X_2^{N_{2k}^{v;e_1,e_2}}$ and $X_1^{N_{2k}^{v;e_1,e_2}}$.

We note that in addition to copies of $X_2$ which are received by node $v$ along $e_1$ at iteration $k$, $v$ may also receive copies of $X_1$ along $e_1$ at iteration $k$. These correspond to closed walks of length $2k$ which start and end at edge $e_1$, and are clearly not cycles. To eliminate these structures in the counting process of $N_{2k}^{v;e_1,e_2}$, one should consider the power of received variables along $e_1$ and $e_2$ excluding the initial message. To describe this, we use the notation $m_{E,v\xleftarrow{e_1}}^{(k)}$ to denote the incoming message to node $v$ along $e_1$ at iteration $k$, excluding the variable of the initial message passed by $v$ along $e_1$. In the above scenario, we have $m^{(k)}_{E,v\overleftarrow{e_1}} = X_2^{N^{v;e_1,e_2}_{2k}}$, and $m^{(k)}_{E,v\overleftarrow{e_2}} = X_1^{N^{v;e_1,e_2}_{2k}}$. This results in

$$N^{v;e_1,e_2}_{2k} = \{\mathrm{ex}(m^{(k)}_{E,v\overleftarrow{e_1}}) + \mathrm{ex}(m^{(k)}_{E,v\overleftarrow{e_2}})\}/2, \tag{1}$$

where $\mathrm{ex}(\cdot)$ is the exponent of the monomial, defined as the sum of the powers of all its variables.

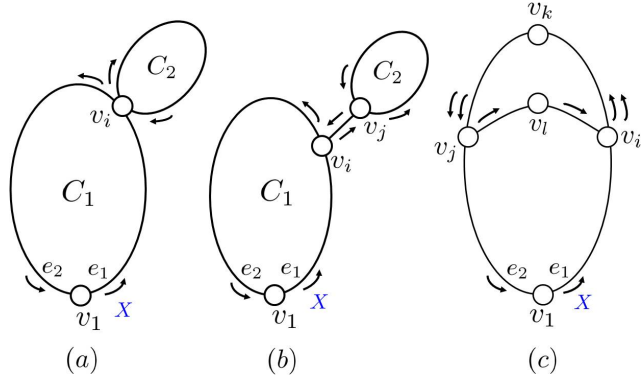

Fig. 3.  Three problematic structures for which the incoming extrinsic messages do not represent cycles.

There is also a possibility that node $v$ receives additional copies of $X_2$ along $e_1$ and additional copies of $X_1$ along $e_2$ at iteration $k$. These additional copies travel either through the same cycle multiple times or through non-cycle lollypop-style closed walks of length $2k$ which start and end at $e_1$ and $e_2$, respectively. Examples of the latter structures are given in Fig. 3, where the message $X$ is initiated at node $v_1$. In Fig. 3(a), $2k$ is in fact the sum of the lengths of the two cycles $C_1$ and $C_2$, while in Fig. 3(b), it is the sum of the lengths of the two cycles plus twice the length of the path between $v_i$ and $v_j$. In Fig. 3(c), message $X$ travels from $v_1$ to $v_i$ first, and then from $v_i$ to $v_j$ through $v_k$. It then travels back from $v_j$ to $v_i$ through $v_l$ followed by a trip from $v_i$ to $v_j$ through $v_k$ for the second time. The journey finally ends when $X$ is passed back from $v_j$ to $v_1$. In this case, the total length of the walk is $2k$.

A careful inspection of the problematic structures, as described above, reveals that they all include at least two cycles. This implies that the shortest length of such structures is $2g$, where $g$ is the girth of the graph. We thus have the following:

*Lemma 2:* Consider a bipartite graph $G = (V, E)$ with girth $g$. Select a node $v \in V$ with two adjacent edges $e_1$ and $e_2$. Assume that the message-passing algorithm is initiated at $t = 0$ by passing 1 along every edge in $E$, except $e_1$ and $e_2$. For $e_1$ and $e_2$ the initial messages are set to monomials $X_1$ and $X_2$, respectively. Then, at iteration $k, k < g/2$, node $v$ will only receive 1 along all its edges including $e_1$ and $e_2$. At iteration $k, g/2 \le k \le g-1$, node $v$ will receive monomials $X_1^i X_2^{N^{v;e_1,e_2}_{2k}}$ and $X_1^{N^{v;e_1,e_2}_{2k}} X_2^j$ along $e_1$ and $e_2$, respectively, where $i$ and $j$ are non-negative integers. Equation (1) is thus valid for $k \le g-1$.

*Proof:* Node $v$ will receive messages other than 1 only if a copy of $X_1$ or $X_2$ is passed back to it. Due to the extrinsic nature of message-passing, such a copy must travel through a lollipop-style closed walk with both ends at $v$. Since the length of a lollipop-style closed walk is at least $g$, no messages other than 1 will be received by $v$ at iterations $k, k < g/2$. At iterations $k, g/2 \leq k \leq g-1$, node $v$ can receive copies of $X_1$ and $X_2$ that have traveled through lollipop-style closed walks with both ends at $v$. In particular, the number of copies of $X_1$ and $X_2$ that $v$ receives at iteration $k \geq g/2$, along $e_2$ and $e_1$, respectively, is equal to the number of lollipop-style closed walks of length $2k$ that start and end at $e_1$ and $e_2$. For $k$ in the range $g/2 \leq k \leq g-1$, such lollipop-style closed walks are limited to cycles of length $2k$ that include $e_1$ and $e_2$. (For $k \geq g$, in addition to cycles, they can include multiple trips over the same cycle or cases such as those in Fig 3.) ∎

Let us now focus on the problem of counting *all* the cycles which pass through a certain node $v$ in a bipartite graph $G = (U \cup W, E)$. Without loss of generality, we assume $v \in U$. One approach to count all the cycles containing $v$ is to use Lemma 2 and count the cycles involving different adjacent edges, two at a time, and then add up the results for any cycle length. The following lemma however suggests a more efficient approach.

*Lemma 3:* Consider a bipartite graph $G = (U \cup W, E)$ with girth $g$, and a node $v \in U$. Initiate the message-passing algorithm by passing 1 on all the edges connected to nodes $u \in U$, $u \neq v$, while passing $d_v$ different monomials, say $X_1, X_2, \ldots, X_{d_v}$, along the edges connected to $v : e_1, \ldots, e_{d_v}$, respectively. For $k \leq g-1$, we then have

$$N_{2k}^v = \sum_{j=1}^{d_v} \operatorname{ex}(m_{E,v \xleftarrow{} e_j}^{(k)})/2 \, , \tag{2}$$

where $N_{2k}^v$ is the number of $2k$-cycles containing $v$.

*Proof:* At iteration $k \leq g-1$, consider the message received by $v$ along $e_j$, $j = 1, \ldots, d_v$, excluding the variable $X_j$. In this extrinsic message $m_{E,v \xleftarrow{} e_j}^{(k)}$, the power of variable $X_i$, $i \neq j$, is $N_{2k}^{v;e_i,e_j}$. We therefore have

$$\operatorname{ex}(m_{E,v \xleftarrow{} e_j}^{(k)}) = \sum_{\substack{i=1 \\ i \neq j}}^{d_v} N_{2k}^{v;e_i,e_j} \, .$$

This combined with

$$N_{2k}^v = \frac{1}{2} \sum_{j=1}^{d_v} \sum_{\substack{i=1 \\ i \neq j}}^{d_v} N_{2k}^{v;e_i,e_j} \, ,$$

completes the proof. ∎

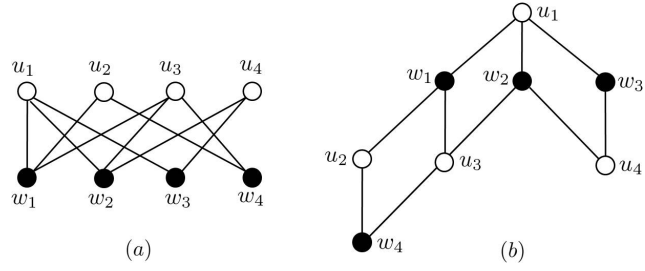

Fig. 4.   Bipartite graph of the example in Section III.C: a) $G$, b) $G$ unwound from node $u_1$.

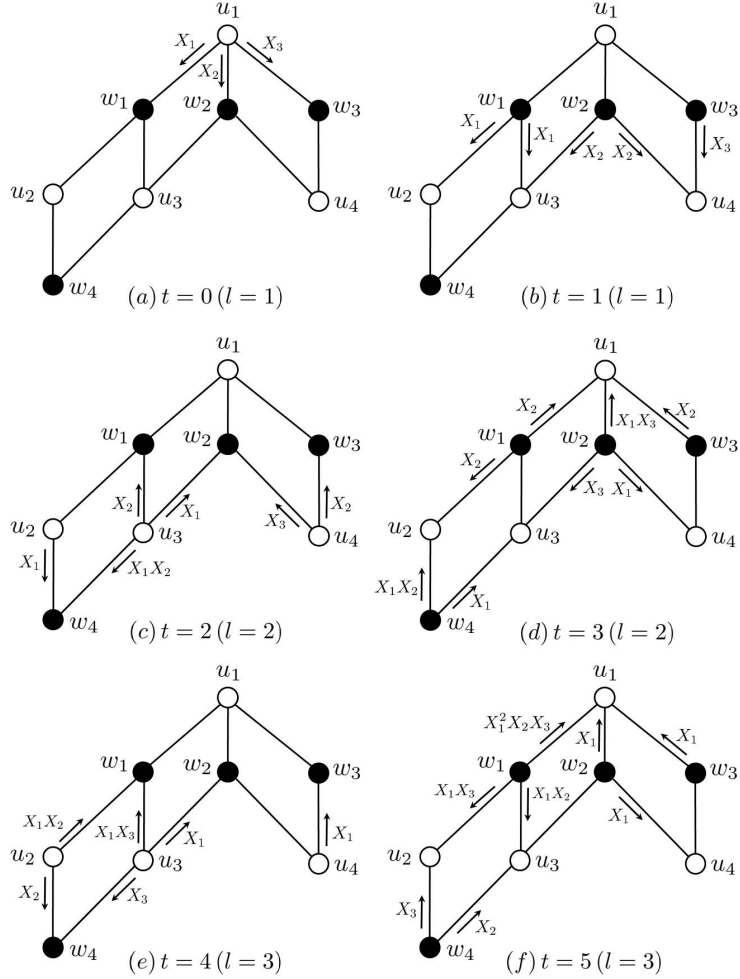

Fig. 5.   Message passing of the proposed algorithm for three iterations in the graph of Fig. 4.

In Lemma 3, at iteration $k, k < g/2$, node $v$ will only receive 1 along all its edges, indicating there are no cycles of length $g - 2$ or smaller containing $v$.

It is worth noting that the message-passing algorithm can be simplified by allowing node $v$ to always pass 1 after the first iteration. This is demonstrated in the following example.

## C. A Simple Example

Here, we illustrate the proposed method by a simple example. Consider the bipartite graph $G$ shown in Fig. 4(a), where the nodes in $U$ and $W$ are represented by hollow and full circles, respectively. Suppose that we are interested in counting short cycles containing node $u_1$. For the simplicity of presentation, as shown in Fig. 4(b), we can unwind the graph G from node $u_1$. It is easy to see from Fig. 4(b) that the girth of $G$ is 4. Using the purposed method, we can thus count cycles of length up to $2g - 2 = 6$. The message-passing algorithm is illustrated in Figures 5(a)-(f):

(a) At $t = 0$, the algorithm is initiated by node $u_1$ passing messages $X_1, X_2$, and $X_3$ along its 3 edges. All the other messages sent by nodes $u_2$, $u_3$ and $u_4$ along their edges are equal to 1, and not shown. [Equivalently, in the vector representation, the initial messages of node $u_1$ are vectors $(1, 0, 0), (0, 1, 0)$ and $(0, 0, 1)$, while all the other messages are $(0, 0, 0)$.]

(b) At $t = 1$, only the nodes in $W$ are active. The corresponding (non-one) messages are shown in Fig 5(b). Note that in this iteration ($\ell = 1$), all the incoming messages to node $u_1$ are equal to one.

(c) At $t = 2$, nodes in $U$ are active. They all pass extrinsic messages using multiplication. For example, $m_{u_3 \to w_4}^{(2)} = m_{w_1 \to u_3}^{(1)} \times m_{w_2 \to u_3}^{(1)} = X_1 X_2$. [In the vector representation, $m_{u_3 \to w_4}^{(2)} = (1, 0, 0) + (0, 1, 0) = (1, 1, 0)$.]

(d) At $t = 3$ ($\ell = 2$), for the first time node $u_1$ receives non-one messages, an indication that there is at least one cycle of length $2\ell = 4$ containing $u_1$. Using (2), we obtain $N_4^{u_1} = (1 + 2 + 1)/2 = 2$.

(e) At $t = 4$, the nodes in $U$ are active and pass messages.

(f) At $t = 5$ ($\ell = 3$), nodes in $W$ are active. Again in this iteration, node $u_1$ receives non-one messages, an indication that it belongs to at least one 6-cycle. Using (2), we have $N_6^{u_1} = (2 + 1 + 1)/2 = 2$.

## IV. Proposed Algorithm

### A. Pseudo Code

To count the short cycles of a certain length $2k$ in the whole graph $G = (U \cup W, E)$, one can apply the proposed algorithm described in the previous section to every node in one of the node partitions, $U$ or $W$, and then add up the results for each cycle length. In this case, for each cycle length, the result should be divided by $k$ as every cycle is counted $k$ times:

$$N_{2k} = (\sum_{u \in U} N_{2k}^u)/k = (\sum_{w \in W} N_{2k}^w)/k \ , \frac{g}{2} \leq k \leq g - 1 \ . \tag{3}$$

To simplify the algorithm and to avoid the k-fold counting repetition, we can deactivate a node as soon as its cycles are counted. This would be equivalent to removing the node and all its adjacent edges from the graph. Moreover, the algorithm can be further simplified by only activating nodes that have at least one non-one incoming message. Based on these simplifications, the proposed algorithm has the pseudo code provided in Algorithm 1.

Algorithm 1 is initiated from $U$. Similarly, it can be initiated from $W$. Nodes in $U$ are indexed by $i = 1, \ldots, n$, and notation $m_{E,(w_j \to u_i)}$ is used to denote the incoming message from node $w_j$ to node $u_i$ excluding the initial variable passed from $u_i$ to $w_j$. Notation $N(u)$ is used for the nodes adjacent to $u$ (neighbors of $u$).

Here we have implicitly assumed that the girth $g$ of the graph is known. In the following subsection, we discuss a modification of the algorithm that can compute $g$ and $N_g$.

### B. Parallel Implementation

The algorithm presented in the previous subsection is based on sequentially going through the nodes in one of the two partitions in the graph. To speed up the counting process and at the expense of larger memory usage, one can run a parallel version of the algorithm in which all the nodes in one partition are initialized simultaneously. This is explained in Fig. 6(a) for the graph of Fig. 4.

The parallel implementation, just described, can also be used to compute $g$ and $N_g$. To see this, note that in the parallel implementation, none of the nodes in the initiating partition will receive a copy of its initial messages before iteration $g/2$. At iteration $g/2$, the nodes which are contained in the shortest cycles will receive copies of their initial messages and all such copies are received along the edges whose initial messages differ from the received messages. This means that all the received copies represent true $g$-cycles. Therefore to compute $g$ and $N_g$, one does not need to distinguish among the initial messages of a node. The initialization in this case is explained in Fig. 6(b) for the graph of Fig. 4. In this setup, if the first iteration in which at least one of the nodes receives a non-one message is iteration $k$, then $g = 2k$, and the number of $g$-cycles is equal to the total number of received non-one messages by all the nodes divided by $2k$.

## V. Complexity of the Proposed Algorithm

### A. Computational Complexity

In the following, we arbitrarily assume that the algorithm is initiated from the node set $U$. We consider a sequential implementation, where the nodes in $U$ are processed one at a time. We also consider the

**Algorithm 1** Proposed Message-Passing Algorithm for Counting Short Cycles

---

**for** $k = 1 : g - 1$ **do**
   $counter(k) = 0$
**end for**
**for** $i = 1 : n$ **do**
  **Initialization**
  $l = 1$
  **for** $w_j \in N(u_i)$ **do**
    $m_{u_i \to w_j}^{(0)} = X_l$
    $l = l + 1$
  **end for**
  **for** $i' = i + 1 : n$ **do**
    **for** $w_j \in N(u_{i'})$ **do**
      $m_{u_{i'} \to w_j}^{(0)} = 1$
    **end for**
  **end for**

  **for** $k = 1 : g - 1$ **do**
    **Message Passing from** $W$
    **for** $j = 1 : m$ **do**
      **for** $u_{i'} \in N(w_j)$ **do**
        $m_{w_j \to u_{i'}}^{(2k-1)} = \prod_{u_h \in N(w_j),\, h \geq i,\, h \neq i'} m_{u_h \to w_j}^{(2k-2)}$
      **end for**
    **end for**

    **Counting Cycles**
    $local\text{-}counter(k) = \sum_{w_j \in N(u_i)} \mathbf{ex}(m_{E,(w_j \to u_i)}^{(2k-1)})$

    **Message Passing from** $U$
    **for** $i' = i + 1 : n$ **do**
      **for** $w_j \in N(u_{i'})$ **do**
        $m_{u_{i'} \to w_j}^{(2k)} = \prod_{w_h \in N(u_{i'}),\, h \neq j} m_{w_h \to u_{i'}}^{(2k-1)}$
      **end for**
    **end for**

  **end for**

  **for** $k = 1 : g - 1$ **do**
    $counter(k) = counter(k) + local\text{-}counter(k)/2$
  **end for**
**end for**

---

vector representation of messages and first derive the complexity for a regular graph. We then generalize the results to irregular graphs. For a regular graph $G = (U \cup W, E)$, starting from a node $u \in U$, there are $d_u$ initial messages, each represented by a unit vector of length $d_u$. All the subsequent messages are also vectors of length $d_u$. To calculate the messages at an active node $w \in W$, we first add all the incoming vectors to $w$, and then subtract from this, the incoming message along each adjacent edge to obtain the outgoing message along that edge. This requires $(2d_w - 1)d_u$ integer additions and subtractions. Similarly, for each active node $u \in U$, we need $(2d_u - 1)d_u$ integer additions and subtractions to obtain the outgoing messages. Considering that in even and odd time instances, the number of active nodes are upper bounded by $n$ and $m$, respectively, the number of operations per iteration is $O(nd_u^2 + md_ud_v) = O(|E|d_u)$. Since the algorithm needs to perform $g - 1$ iterations, the complexity of the algorithm for each node $u \in U$ is $O(gnd_u^2 + gmd_ud_v) = O(g|E|d_u)$. The total complexity is thus

$$O(gn^2d_u^2 + gnmd_ud_v) = O(gn^2d_u^2) = O(g|E|^2) .$$

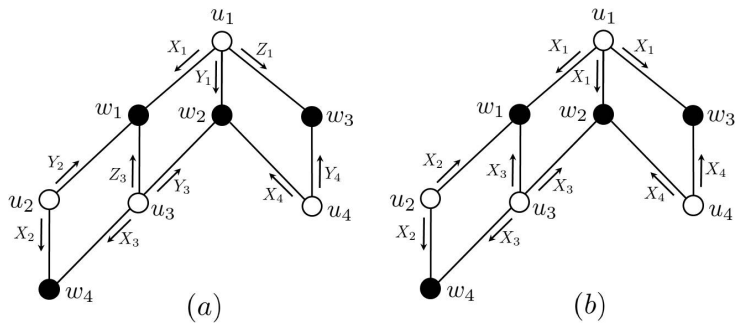

Fig. 6. Initial message-passing in parallel implementations: a) counting short cycles of length $2k$, $g/2 \leq k \leq g - 1$, b) calculating $g$ and $N_g$.

It is easy to see that the same complexity order also applies to irregular bipartite graphs.

In the above discussions, it is implicitly assumed that the girth of the graph is known a priori. Since the computational complexity of finding the girth is at most $O(n^2)$, e.g., based on the algorithm of [11],[1] the extra complexity for computing the girth is negligible compared to the rest of the computations.

### B. Memory Requirements

For each edge of the bipartite graph, we need two memory locations to store the message vectors in both directions. For a regular graph, since each vector has $d_u$ elements, the total number of memory locations, each storing an integer number, is $2d_u|E|$ or $O(nd_u^2) = O(d_u|E|)$. For an irregular graph, the storage complexity is $O(d_{max}|E|)$, where $d_{max}$ is the maximum node degree in $U$ or $W$, depending on which side initiates the algorithm.

---

[1]It is easy to see that if we use the algorithm proposed in Section IV.B to compute $g$, the complexity is $O(gn^2d_u)$, which is in general larger than that of [11]. The algorithm of [11] however only finds $g$, while the proposed algorithm also computes $N_g$.

TABLE I
NUMBER OF SHORT CYCLES IN THE TANNER GRAPHS OF FOUR RATE-1/2 LDPC CODES

|          | Code A   | Code B | Code C | Code D |
|----------|----------|--------|--------|--------|
| $N_6$    | 11538    | 0      | 179    | 161    |
| $N_8$    | 408657   | 2      | 1218   | 1260   |
| $N_{10}$ | 13110235 | 11238  | 9989   | 10051  |
| $N_{12}$ | -        | 91101  | -      | -      |
| $N_{14}$ | -        | 748343 | -      | -      |

### C. Comparison with the Algorithm of [8]

First, it is important to note that while the algorithm of [8] is limited to bipartite graphs, the proposed algorithm is capable of counting short cycles in a general (non-bipartite) graph. For bipartite graphs, the algorithm of [8] counts cycles of length $g, g + 2, g + 4$, while the proposed algorithm counts cycles of length $g, g+2, \ldots, 2g-2$. The coverage of the proposed algorithm is thus at least as much as the algorithm of [8] for graphs with $g \geq 6$. It should be noted that the Tanner graphs of almost all good LDPC codes have $g \geq 6$.

The computational complexity of the algorithm of [8] is $O(gn^3)$, where $n = \max(|U|, |W|)$. The complexity of the proposed algorithm is $O(g|E|^2)$. One can thus see that for sparse graphs with $|E|$ growing slower than $n^{3/2}$, the complexity of the proposed algorithm is less than that of the algorithm in [8]. Moreover the computations in the algorithm presented here are simple integer additions and subtractions, while in [8] the operations are mainly high-precision multiplications.

In terms of memory requirements, the algorithm of [8] requires at most $11(n^2 + m^2) + 21nm$ high bit-width (64-bit integer) storage locations, which is of order $O(n^2)$. The proposed algorithm on the other hand requires $2d_u|E|$ memory locations, i.e., $O(d_{max}|E|)$, which for sparse graphs can be much smaller than what is needed for the algorithm of [8]. Moreover, the maximum size of memory locations for the proposed algorithm, which is proportional to the number of cycles, is usually much less than 64 bits.

## VI. NUMERICAL RESULTS

In this section, we present numerical results obtained by applying the proposed algorithm to Tanner graphs of LDPC codes. We consider four rate-1/2 codes from [16]. Codes A and B are listed in [16] as $PEGirReg504x1008$ and $PEGReg504x1008$, respectively. Both codes are constructed using the Progressive Edge Growth (PEG) method of [9], and have $n = 1008$ and $m = 504$. Code A is irregular while Code B is regular. Codes C and D are MacKay's codes $8000.4000.3.483$ and $10000.10000.3.631$, respectively. They are both regular with $d_u = 3$ and $d_w = 6$. For Code C, $n = 8000$ and $m = 4000$, while these parameters for Code D are $20,000$ and $10,000$, respectively. The number of short cycles in the Tanner graphs of these codes is listed in Table I. Codes A, C and D have girth 6 and the proposed algorithm, similar to the algorithm of [8], can compute $N_6$, $N_8$ and $N_{10}$. Code B however has girth 8, and while the algorithm of [8] can only compute $N_8$, $N_{10}$ and $N_{12}$, the proposed algorithm can also compute $N_{14}$.

TABLE II
CPU TIME AND MEMORY REQUIREMENTS FOR THE PROPOSED ALGORITHM

|  | CPU Time (S) | Max Memory (MB) | Max Swap (MB) |
|---|---|---|---|
| Code A | 5.3 | 0.36 | 3.3 |
| Code B | 3 | 0.36 | 2.8 |
| Code C | 155 | 13 | 157 |
| Code D | 1127 | 13 | 157 |

TABLE III
CPU TIME AND MEMORY REQUIREMENTS FOR THE ALGORITHM OF [8]

|  | CPU Time (S) | Max Memory (MB) | Max Swap (MB) |
|---|---|---|---|
| Code A | 10.3 | 1.5 | 35 |
| Code B | 16.6 | 1.5 | 35 |
| Code C | 4965 | 7839 | 14195 |
| Code D | - | - | - |

Tables II and III show the running time and memory requirements of the proposed algorithm and the algorithm of [8],[2] respectively. Both algorithms were run on the same machine with a 2.2-GHz CPU and 8 GB of RAM. As can be seen, the proposed algorithm is consistently faster than the algorithm of [8] and requires significantly less memory for larger graphs. In fact, for Code D, the algorithm of [8] ran out of memory and was not able to find the results.

As another experiment, we randomly generate six parity-check matrices for each of the following three rate-1/2 LDPC code ensembles: $(d_u, d_w) = (3, 6), (4, 8), (5, 10)$. The lengths for each degree distribution are: $n = 200, 500, 1000, 5000, 10,000$ and $20,000$. In the generation of the parity-check matrices, 4-cycles are avoided. The proposed algorithm is then used to count the short cycles of each parity-check matrix. The results, which are reported in Table IV, show that while there is a large difference between the short cycle distribution of different degree distributions, the changes with respect to the block length for the same degree distribution are negligible. This would imply that the complexity of the algorithms which are

---

[2]To implement the algorithm of [8], we used the authors' code in [17].

TABLE IV
DISTRIBUTION OF SHORT CYCLES IN THE TANNER GRAPHS OF RATE-1/2 RANDOM REGULAR LDPC CODES WITH
DIFFERENT DEGREE DISTRIBUTIONS AND DIFFERENT BLOCK LENGTHS

| Degree Distribution | Short Cycle Distribution | Code Lengths | | | | | |
|---|---|---|---|---|---|---|---|
| | | 200 | 500 | 1000 | 5000 | 10000 | 20000 |
| $(3, 6)$ | $N_6$ | 171 | 167 | 181 | 156 | 166 | 148 |
| | $N_8$ | 1265 | 1239 | 1226 | 1235 | 1253 | 1285 |
| | $N_{10}$ | 10069 | 10110 | 9939 | 9982 | 9858 | 9974 |
| $(4, 8)$ | $N_6$ | 1636 | 1611 | 1584 | 1562 | 1537 | 1572 |
| | $N_8$ | 25005 | 24419 | 24379 | 24363 | 24529 | 24557 |
| | $N_{10}$ | 409335 | 409373 | 408595 | 407958 | 408246 | 409051 |
| $(5, 10)$ | $N_6$ | 8626 | 8064 | 8055 | 7978 | 7858 | 7926 |
| | $N_8$ | 213639 | 212484 | 210767 | 210153 | 209614 | 210159 |
| | $N_{10}$ | 6052158 | 6054661 | 6049148 | 6043400 | 6049583 | 6053704 |

based on the enumeration of short cycles in a Tanner graph is rather independent of the block length.[3]

# VII. CONCLUSION

In this paper, we proposed a distributed message-passing algorithm to count short cycles in a graph. For bipartite graphs, the proposed algorithm counts short cycles of length $g, g + 2, \ldots, 2g - 2$, where $g$ is the girth of the graph. For non-bipartite graphs, the algorithm counts cycles of length $g, g + 1, \ldots, 2g - 1$. The operations performed by the algorithm are integer additions and subtractions, and the computational and storage complexities of the algorithm are $O(g|E|^2)$ and $O(d_{max}|E|)$, respectively, where $|E|$ and $d_{max}$ are the number of edges and the maximum node degree in the graph, respectively. For sparse graphs, the proposed algorithm is significantly faster and requires substantially less memory compared to the existing algorithms, particularly for larger graphs.

# REFERENCES

[1] N. Alon, R. Yuster and U. Zwick, "Finding and counting given length cycles," *Algorithmica*, vol. 17 no. 3, pp. 209-223, 1997.

[2] Reza Asvadi, A. H. Banihashemi and M. Ahmadian-Attari, "Lowering the Error Floor of LDPC Codes Using Cyclic Liftings," accepted for presentation at the *2010 Int. Symp. Inform. Theory (ISIT)*, Austin, Texas, June 13 - 18, 2010. Available at http://arxiv.org/PS_cache/arxiv/pdf/1002/1002.4311v1.pdf.

[3] C. Berrou, A. Glavieux, and P. Thitimajshima, "Near Shannon limit errorcorrecting coding and decoding: Turbo-codes," in Proc. *IEEE Int. Conf. Comm.*, Geneva, Switzerland, pp. 1064-1070, May 1993, .

[4] R. Chen, H. Huang and G. Xiao, "Relation between parity-check matrices and cycles of associated Tanner graphs," *IEEE Comm. Lett.*, vol. 11, no. 8, pp. 674 - 676, Aug. 2007.

---

[3]It is worth mentioning that it is proved in [12] that for random regular bipartite graphs with $d_u = d_w = d$, as the number of nodes tends to infinity, the distribution of short cycles of different length $c_i$ tends to independent Poisson distributions with average $\mu_i = (d - 1)^{c_i}/c_i$. To the best of our knowledge, however, no generalization of this result is available for bipartite graphs with $d_u \neq d_w$.

[5] J. Fan and Y. Xiao, "A method of counting the number of cycles in LDPC codes," *8th Int. Conf. Signal Proc. (ICSP)*, vol. 3, pp. 2183-2186, 2006.

[6] J. Flum and M. Grohe, "The parameterized complexity of counting problems," in Proc. *IEEE Symp. Foundations of Computer Science*, Vancouver, BC, Canada, pp. 538-547, Nov. 2002.

[7] R. G. Gallager, "Low density parity check codes," *IEEE Trans. Inform. Theory*, vol. 8, no. 1, pp. 21-28, Jan. 1962

[8] T.R. Halford and K.M. Chugg, "An algorithm for counting short cycles in bipartite graphs," *IEEE Trans. Inform. Theory*, vol. 52, no. 1, pp. 287-292, Jan. 2006.

[9] X.-Y. Hu, E. Eleftheriou, and D. M. Arnold, "Regular and irregular progressive edge-growth Tanner graphs," *IEEE Trans. Inform. Theory*, vol. 51, no. 1, pp. 386-398, Jan. 2005.

[10] F. R. Kschischang , B. J. Frey and H.-A. Loeliger "Factor graphs and the sum-product algorithm," *IEEE Trans. Inform. Theory*, vol. 47, no. 2, pp. 498-519, Feb. 2001.

[11] Y. Mao and A. H. Banihashemi, "A heuristic search for good low-density parity-check codes at short block lengths," in Proc. *IEEE Int. Conf. Comm.*, vol. 1, Helsinki, Finland, pp. 41-44, June 2001.

[12] B. D. McKay, N. C. Wormald and B. Wysocka, "Short cycles in random regular graphs," *Electronic Journal of Combinatorics*, 11(1):#R66, 2004.

[13] T. J. Richardson and R. Urbanke, "The capacity of low-density paritycheck codes under message-passing decoding," *IEEE Trans. Inform. Theory*, vol. 47, no. 2, pp. 599 - 618, Feb. 2001.

[14] R. M. Tanner, A recursive approach to low-complexity codes, *IEEE Trans. Inform. Theory*, vol. 27, pp. 533-547, 1981

[15] H. Xiao and A. H. Banihashemi, "Error Rate Estimation of Low-Density Parity-Check Codes on Binary Symmetric Channels Using Cycle Enumeration," *IEEE Trans. Comm.*, vol. 57, no. 6, pp. 1550 - 1555, June 2009.

[16] [Online]. Available: http://www.inference.phy.cam.ac.uk/mackay/codes/ data.html

[17] [Online]. Available: http://csi.usc.edu/chugg/tools